%% file: otic_blueprint.tex
\documentclass[conference, nofonttune]{IEEEtran}
\IEEEoverridecommandlockouts
\usepackage{cite}
\usepackage{amsmath,amssymb,amsfonts}
\usepackage{algorithmic}
\usepackage{graphicx}
\usepackage{textcomp}
\usepackage{xcolor}
\usepackage{todonotes}
\usepackage{booktabs}
\usepackage{amsmath}
\usepackage{gensymb}
\usepackage{multirow, makecell}
\usepackage{enumitem}
\def\BibTeX{{\rm B\kern-.05em{\sc i\kern-.025em b}\kern-.08em
    T\kern-.1667em\lower.7ex\hbox{E}\kern-.125emX}}

\usepackage{soul}
\usepackage[font=footnotesize]{caption}
\usepackage[position=b,font=footnotesize]{subcaption}
\usepackage{xspace}
\usepackage{url}

\newcommand*{\duts}{\glspl{dut}\xspace}
\newcommand*{\cicd}{\gls{ci}/\gls{cd}\xspace}
\newcommand*{\ct}{\gls{ct}\xspace}

\usepackage[acronyms,nonumberlist,nopostdot,nomain,nogroupskip,acronymlists={hidden}]{glossaries}
\newglossary[algh]{hidden}{acrh}{acnh}{Hidden Acronyms}
\glsdisablehyper
\input{acronyms.tex}

\begin{document}
\bstctlcite{IEEEexample:BSTcontrol}
\title{Open6G OTIC: A Blueprint for Programmable O-RAN and 3GPP Testing Infrastructure\vspace{-.45cm}
\thanks{This article is based upon work partially supported by the O-RAN ALLIANCE, by the National Telecommunications and Information Administration (NTIA)'s Public Wireless Supply Chain Innovation Fund (PWSCIF) under Award No. 25-60-IF054, and by the Massachusetts Technology Collaborative under award 22563. The authors would also like to thank collaborators at Keysight Technologies and VIAVI Solutions for the valuable feedback.}
}


\author{
    \IEEEauthorblockN{Gabriele Gemmi, Michele Polese, Pedram Johari, Stefano Maxenti, Michael Seltser\IEEEauthorrefmark{1}, Tommaso Melodia}
    \IEEEauthorblockA{Institute for the Wireless Internet of Things, Northeastern University, Boston, MA}
    \IEEEauthorblockA{\IEEEauthorrefmark{1}Cerbo IO, Burlington, MA}
    \IEEEauthorblockA{open6g.otic@northeastern.edu}
}

\makeatletter
\patchcmd{\@maketitle}
  {\addvspace{0.5\baselineskip}\egroup} 
  {\addvspace{-1\baselineskip}\egroup} 
  {}
  {}
\makeatother

\maketitle

\begin{abstract}
Softwarized and programmable \glspl{ran} come with virtualized and disaggregated components, increasing the supply chain robustness and the flexibility and dynamism of the network deployments. This is a key tenet of Open \gls{ran}, with open interfaces across disaggregated components specified by the O-RAN ALLIANCE. It is mandatory, however, to validate that all components are compliant with the specifications and can successfully interoperate, without performance gaps with traditional, monolithic appliances. \glspl{otic} are entities that can verify such interoperability and adherence to the standard through rigorous testing. However, how to design, instrument, and deploy an \gls{otic} which can offer testing for multiple tenants, heterogeneous devices, and is ready to support automated testing is still an open challenge. In this paper, we introduce a blueprint for a programmable \gls{otic} testing infrastructure, based on the design and deployment of the Open6G OTIC at Northeastern University, Boston, and provide insights on technical challenges and solutions for O-RAN testing at scale.
\end{abstract}

\begin{IEEEkeywords}
Open RAN, OTIC, 5G, Testing, Infrastructure
\end{IEEEkeywords}

\section{Introduction}


\begin{figure*}[h]
  \centering
  \hfill
  \begin{subfigure}[t]{0.28\linewidth}
    \includegraphics[width=.8\textwidth]{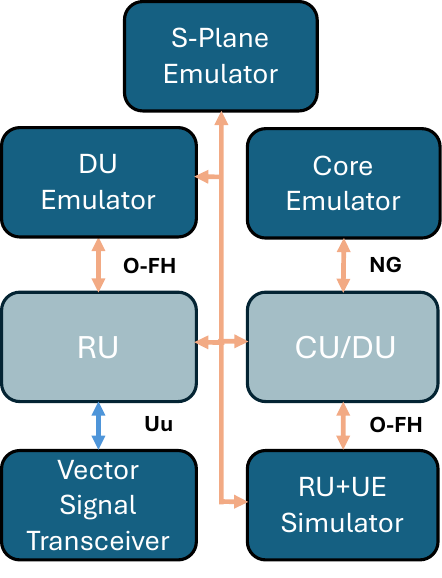}
    \caption{WG4 Conformance}
    \label{fig:wg4_conformance}
  \end{subfigure}
  \hfill
    \begin{subfigure}[t]{0.28\linewidth}
    \includegraphics[width=.9\textwidth]{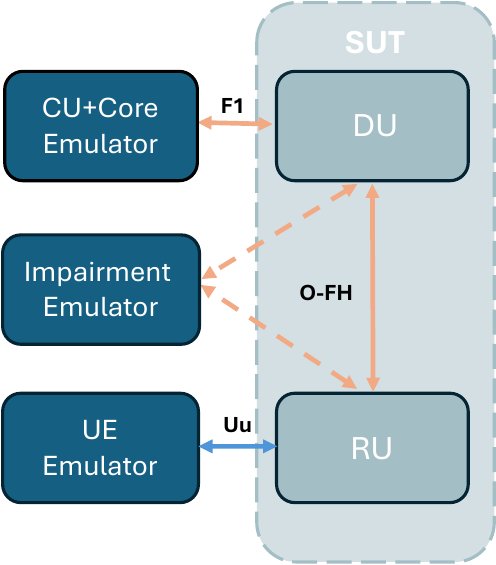}
    \caption{WG4 \gls{iot}}
    \label{fig:wg4_iot}
  \end{subfigure}
  \hfill
  \begin{subfigure}[t]{0.28\linewidth}
    \includegraphics[width=.9\textwidth]{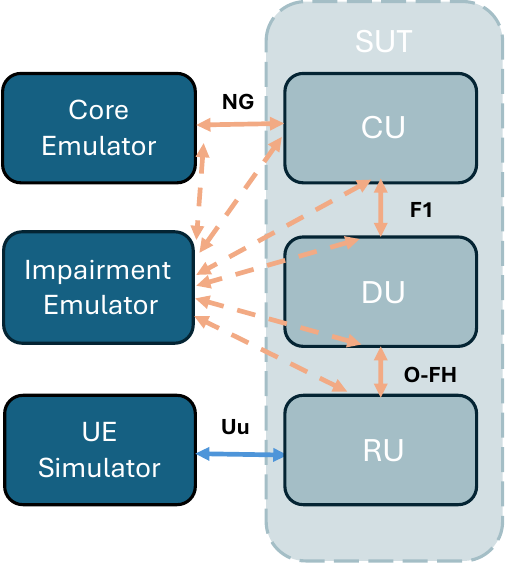}
    \caption{TIFG End to End (E2E)}
    \label{fig:e2e}
  \end{subfigure}
  \hfill
  \caption{Different \gls{wat} topologies for Open RAN. Orange arrows represent digital communication, while blue arrows represent analog (RF) communication.}
  \vspace{-.45cm}
  \label{fig:testing_topologies}
\end{figure*}

Network disaggregation, softwarization, and multi-vendor deployments are foundational principles of Open \gls{ran}~\cite{polese2023understanding}, providing much-needed market diversity and supply chain robustness~\cite{dellOroRAN}. Softwarization also enables programmable protocol stacks that can be optimized through closed-loop control exercised by \glspl{ric}~\cite{bala2021ric}. At the same time, these ingredients also introduce interoperability, performance, and security challenges~\cite{groen2024securing,klement2024securing}. Reports indicate that interoperability and performance testing can de-risk Open RAN deployments and pave the way to more flexible networks~\cite{dellOroRAN}, as well as programmable networks with performance on par or better than that of monolithic networks~\cite{garciaaviles2021nuberu}.

Overall, network disaggregation is increasing the complexity of the wireless supply chain, as a more diverse set of devices and open interfaces need validation against various specifications from the O-RAN ALLIANCE and 3GPP~\cite{polese2023understanding}. This complexity is exacerbated by the rapid growth of private cellular networks~\cite{cagr_private}, which involve new entrants like system integrators and startups that lack established testing facilities. Although network disaggregation and softwarization could reduce costs for these networks, insufficient or costly testing processes may lead these new entrants to rely on traditional vendors, undermining the advantages of open radio systems~\cite{brown2023heavy}.

To summarize, testing for 5G-and-beyond systems is required to unlock supply chain diversity and drive innovation in public/private networks. In this context, the O-RAN ALLIANCE has recognized a number of testing centers as \glspl{otic}, with the mission of facilitating the testing and certification of multi-vendor \gls{ran} products and software~\cite{oranWebsiteOticPage}. 
%
%
As of today, however, testing wireless systems remains a labor-intensive, highly manual effort. Indeed, while \cicd and automation are widely used in cloud systems, existing techniques cannot be directly applied to  cellular solutions, which come with heterogeneous devices, spectrum and radio requirements, distributed deployments, need for high performance and real-time processing, and complex technical specifications to parse and comply with. In addition, tests can concern \duts with very different requirements, e.g., \glspl{ru} with fronthaul and \gls{rf} interfaces vs. software for the \gls{cu} or hardware acceleration for the \gls{du}. Finally, testing requires dedicated equipment which is often challenging to provision and operate.

To address these challenges, in this paper we discuss how we designed and deployed a programmable \gls{otic} infrastructure in the Northeastern University Open6G \gls{otic} in Boston, comprising testing equipment, plug-and-play capabilities for \duts, and a cloud environment for the management of tenants and tests. We design a testing infrastructure that (i) supports \gls{otic} operations and a wide set of tests; (ii) is resilient; (iii) allows for the coexistence of multiple tenants with isolation, security, and remote access, and (iv) enables different classes of tests without the need to reconfigure the physical setup. We achieve this by designing and deploying a programmable physical infrastructure that can be instrumented and configured through \glspl{api} and a set of services that support the \gls{otic} requirements.

This novel Open6G OTIC blueprint can inform and guide the development of testing infrastructure for cellular systems. It also represents a decisive step toward automating testing, as \cicd for \ct require an underlying infrastructure that can be instrumented programmatically to meet the configuration that can enable the specific test to be performed. The rest of the paper is organized as follows. Section~\ref{sec:o-ran-testing} reviews the O-RAN testing methodologies. Section~\ref{sec:infra-requirements} maps them into requirements that the \gls{otic} infrastructure should support. Section~\ref{sec:blueprint} describes the architectural blueprint that meets such requirements and is the foundation of the Open6G OTIC. Section~\ref{sec:conclusions} concludes the paper.

\section{Testing Disaggregated Cellular Systems}
\label{sec:o-ran-testing}

\glsreset{wat}
\glsreset{iot}

\glspl{otic} are specialized facilities endorsed by the O-RAN ALLIANCE that provides a collaborative, vendor-neutral environment to support development and validation of Open RAN technologies~\cite{oranWebsiteOticPage}. 
These centers, including the Northeastern University Open6G OTIC, support the O-RAN ecosystem by offering platforms for testing, integration, and certification of O-RAN products and solutions.
Primarily, an \gls{otic} can conduct conformance, \gls{iot} and \gls{e2e} tests to ensure that products meet O-RAN specifications~\cite{oran23overview}.
For instance, an \gls{otic} can evaluate whether an \gls{ru} and its implementation of \gls{ofh} comply with the relevant specifications (e.g.,~\cite{oran23wg4conf}).
Alternatively, one or more vendors could call for the testing of a disaggregated \gls{gnb}, with \gls{cu}, \gls{du}, and \gls{ru}, so that they can be evaluated for interoperability or as a single system for their \gls{e2e} functionalities. 

In this context, \glspl{otic} are also responsible for awarding O-RAN certificates and badges, after a product has passed specific conformance, \gls{iot} or \gls{e2e} tests.
These certifications help vendors demonstrate to the operators the reliability and compatibility of their products, facilitating broader adoption and deployment of Open RAN technologies \cite{oran23overview, oran23cgotic}.
Finally, \glspl{otic} also host O-RAN PlugFests, where multiple vendors interoperate their solutions in real-world scenarios.

All the tests specifications share the same methodology, known as the \gls{wat}. With this, the \gls{dut} is wrapped with \gls{te}, which emulates communication with the other components through open interfaces \cite{oran23wg4conf}.
This technique is also applied to test multiple devices collectively referred to as \gls{sut}. Additional testing appliances such as impairment emulators might be integrated into the \gls{sut} \cite{oran23wg4iot,oran23tifge2e}.
The following paragraphs describe the three main categories of OTIC tests.

\vspace{0.5em}
\noindent\textbf{Conformance Testing} validates the adherence of a single \gls{dut} to the specification for a specific O-RAN interface, facilitating interoperability with devices from different vendors (e.g., DUs and RUs). This process involves wrapping the \gls{dut} with \gls{te}, stimulating it on one interface, and monitoring the output on the other interfaces or logs within the \gls{dut} itself.
Currently, conformance tests has been defined for the \gls{ofh} interface, as specified by \gls{wg} 4~\cite{oran23wg4conf}.
\gls{ofh} conformance is further subdivided in \gls{du} and \gls{ru} conformance tests, indicating which side of the \gls{ofh} interface is being tested.
Fig.~\ref{fig:wg4_conformance} shows the topology and \gls{te} needed for WG4 Conformance for both DU and RU.
While the testing focuses on \gls{ofh}, \gls{te} is also connected to the other interfaces (NG interface with the Core, or the Uu Air Interface with the \gls{ue}) to determine whether the \gls{dut} behaves according to the specifications. 

Conformance testing generally proceeds across four planes. The \textbf{Management Plane (M-Plane)} is tested first to ensure that the \gls{dut} can be managed remotely using a mix of TLS, SSH, and NETCONF protocols. This is crucial for subsequent tests, as the M-Plane is used to automate parts of the tests. The \textbf{Synchronization Plane (S-Plane)} is typically tested next since its correct operation is necessary for testing the remaining planes. This involves using an S-Plane Emulator that generates PTP and SyncE messages representing different values for the synchronization accuracy. The emulator is often integrated into other \gls{te}, such as the \gls{du} emulator. The final tests focus on \textbf{Control and User Planes (CU-Plane)}. For RU conformance testing, a \gls{du} Emulator generates and transmits \gls{ofh} packets representing a specific 5G waveform to the \gls{ru}, which converts them into analog signals transmitted to a \gls{vst} either through antennas (radiated testing) or via coaxial cables (conducted testing). In the VST the signal is digitized and compared with the original one for different metrics. The reverse process is applied to test the uplink. 
For \gls{du} conformance testing, a higher-level test is conducted using an \gls{ru} and \gls{ue} Emulator. This setup transmits \gls{ofh} packets representing signals from emulated \glspl{ue} to the \gls{du}, which is also connected to a Core Emulator.

\vspace{0.5em}
\noindent \textbf{Interoperability Testing} shares the same objective of conformance testing, i.e., to guarantee interoperability across multiple vendors. However, instead of testing a specific device in isolation, it validates that two specific \glspl{dut} can work together seamlessly. 
This type of testing is defined across multiple O-RAN \glspl{wg}, and currently includes specifications from WG4 for the \gls{ofh} interface~\cite{oran23wg4iot} and WG5 for the 3GPP F1, E1, X2, and Xn interfaces \cite{oran23wg5iot}.
Fig.~\ref{fig:wg4_iot} shows the topology and \gls{te} needed for WG4 IOT among a \gls{ru} and a \gls{du}.
In this testing configuration, the two \gls{dut} (collectively called \gls{sut}) are wrapped by a CU and Core Emulator and a UE Emulator.
An Impairment Emulator is also used in delay management test-cases to artificially increase the delay on the fronthaul interface, simulating large distances between the \gls{du} and \gls{ru}.

The first step in DU/RU \gls{iot} testing is drafting the \gls{iot} profile, using a template published by the O-RAN ALLIANCE.
On such document, both the \gls{du} and \gls{ru} vendors will note the features that their devices support, and then a common configuration will be chosen to perform the \gls{iot} tests.
Then, similarly to conformance, \gls{iot} testing proceeds across the four different planes. 
The M-Plane is tested first to ensure that the two \glspl{dut} can leverage the M-Plane to communicate and configure each other automatically, followed by the S-Plane. 
Depending on the Topology the S-Plane might be generated by an external T-GM or by either \gls{dut}. 
Then, the CU-Plane is tested first by validating that \gls{ofh} delay is managed properly, as explained above.
Finally, best-effort performance tests are performed, verifying that data transfers can be executed from the \gls{ue} simulator to the core simulator over the \gls{sut}.

\vspace{0.5em}
\noindent \textbf{End-to-end Testing} is not specific to a single interface or \gls{dut}, but focuses on testing a whole system from a functional and performance points of view \cite{oran23tifge2e}.
\gls{e2e} testing specifications are under the purview of the \gls{tifg} rather than a \gls{wg}, and it involves wrapping a complete disaggregated \gls{gnb} (\gls{cu}, \gls{du}, \gls{ru}) between a \gls{ue} simulator and a Core Simulator. This setup tests all functionalities of a 5G RAN, including basic functionalities and performance, ensuring comprehensive system validation.

Fig.~\ref{fig:e2e} shows the topology and \gls{te} needed to perform \gls{e2e} test cases. The \gls{te} is the same as for \gls{iot} testing, however in this case the Impairment Emulator is also used on F1 and NG interfaces to assess how latency affect the \gls{e2e} functionality  of the system.
Additionally, for mobility test cases multiple \gls{ru} and or \gls{du} might be used to validate handover procedures.

Finally, as testing centers, OTICs may also perform additional tests, e.g., for 3GPP or FCC certification.

\section{Requirement Analysis}
\label{sec:infra-requirements}

As part of the design phase of the Open6G OTIC blueprint, we have identified key requirements that the infrastructure needs to support toward enabling the tests \emph{and} providing support for automated continuous testing:

\begin{itemize}[noitemsep,topsep=0pt,parsep=0pt,partopsep=0pt,leftmargin=*]
    \item \textbf{R1: Enable Zero-Touch Reconfigurability.} As described and shown in the previous section, there are different testing topologies that need to be deployed in order to run the various O-RAN tests, sharing most of the testing appliances. Moreover, there might be multiple DUT of the same kind undergoing the same tests, or even permanently installed and enrolled in a \ct pipeline.
    For this reasons, the first and main requirements is to be able to reconfigure on the fly how these appliances interconnect with each other and with the \glspl{dut}.

    \item \textbf{R2: Isolation Among Tenants.}
    \glspl{otic} are shared infrastructure where multiple tenants test their devices and possibly interoperate with each other. 
    For this reason is imperative for the infrastructure to be properly segmented, so that each tenant cannot get access to other tenants network space, inadvertently expose sensitive information, or possibly compromise the systems of other tenants.

    \item \textbf{R3: Support for Virtualized Applications.}
    Most of the software \gls{te} as well as some \gls{dut} needs to be deployed either as virtual machines or as containers. 
    For this reason the infrastructure needs to be able to accommodate such requirements. 
    Additionally, this infrastructure can be used by the OTIC team to deploy internal services.

    \item \textbf{R4: Resiliency.}
    Once installed, \glspl{te} and \glspl{dut} needs to be always operational.
    Further, this equipment is sensitive to power and temperature fluctuations and thus proper cooling and power stabilization must be provided.

    \item \textbf{R5: Secure Remote Access.} Each tenant needs to access to their own virtual and physical assets so that they can be managed remotely. 
    Moreover, during the testing phases, technical support and R\&D teams will have to connect to the devices to troubleshoot issues and collect debug logs.

    \item \textbf{R6: Seamless Support for Multiple Sites.}
    The \gls{otic} infrastructure might need to span multiple buildings or even sites.
    This might be needed to connect a \gls{ru} installed outdoor for field testing (part of the \gls{e2e} tests), or a \gls{ru} and \gls{vst} installed in an anechoic chamber (for radiated conformance tests).
    
    \item \textbf{R7: Integration with Other Testbeds for \gls{e2e} Testing.}
    \gls{otic} are often hosted in research institutions where other testbeds are usually available. In the case of the Open6G \gls{otic}, X5G, a multi-vendor private 5G Network \cite{villa2024x5gopenprogrammablemultivendor}, and Colosseum~\cite{polese2024colosseumopenrandigital} are also available.
    The \gls{otic} infrastructure benefits from a tight integration with such testbeds, so that customers can leverage these unique facilities.

    \item \textbf{R8: Need for dedicated RAN hardware.}
    Softwarized RAN applications, as well as \gls{te}, often require specialized hardware, such as HW accelerators or 
    Telecom Clocks (T-GMs or T-BCs). 
    The infrastructure deployed need to be able to accommodate for these needs.
    
\end{itemize}

\section{Programmable Architecture Blueprint}
\label{sec:blueprint}


In this section, we will discuss how we designed the OTIC infrastructure to meet the eight requirements discussed above, following a bottom-up approach. 
First we detail the physical deployment and interconnections of the different appliances. We then describe how we can leverage \glspl{vlan} and programmable switches to implement a programmable logical topology. Next, we describe  how the network is addressed and what policies have been put in place to ensure isolation among tenants. Finally, we describe OTIC specific services that complement the infrastructure to facilitate day-to-day operations.

\begin{figure}
    \centering
    \includegraphics[width=1\linewidth]{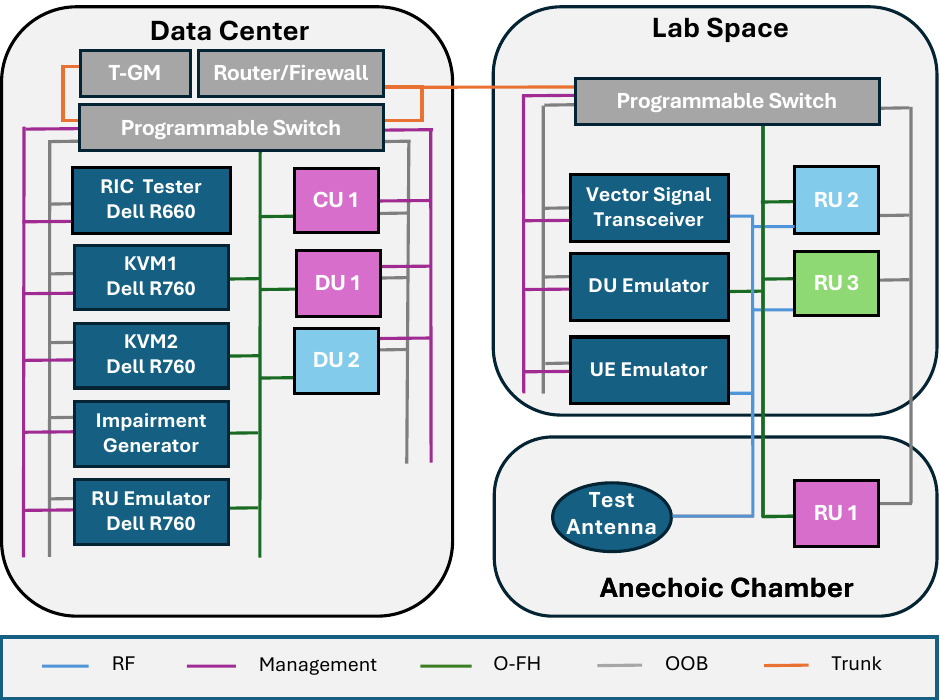}
    \caption{Physical Topology of the Open6G \gls{otic}. Network connections for \gls{oob} are coloured in gray, management in purple, \gls{ofh} in dark green, and trunks between switches/routers are in orange. 
    Analog RF is in blue.} 
    \label{fig:phy_topo}
    \vspace{-.3cm}
\end{figure}

\noindent\textbf{Physical Topology:}
%
%
As shown in Fig. \ref{fig:phy_topo}, the deployment of the appliances in the \gls{otic} is divided in two main areas. 
A data center houses all the purely digital components, such as \glspl{du} from the customers and digital \gls{te} (RU Emulators, Core Simulators, impairment generators, etc). We leverage Dell R760 compute to host most of the software appliances in a KVM cluster, to address requirement {\bf R3}. 
As separate space, such a lab space adjacent to an anechoic chamber, hosts all the analog/digital components, such as \glspl{ru} from the customers and analog \gls{te} (UE Emulators, \glspl{vst}, etc).
To meet {\bf R4}, equipment hosted in the data center is kept at a constant temperature of 20$\degree$ Celsius and is powered through an \gls{ups} unit to ensure a stable and resilient power supply. The data center also co-hosts the Colosseum infrastructure.

In each space, one or multiple programmable switches are used to physically interconnect all the \glspl{te} and \glspl{dut} using the Ethernet protocol, to support requirement {\bf R1}. Specifically, in the Open6G OTIC, we use Dell S5232F-ON and Dell S5248F-ON switches
To address {\bf R6}, the spaces are then interconnected with each other through one or multiple of single-mode fibers together with QSFP28 laser transceivers rated each for 100GbE. 

To address {\bf R8}, we leverage Telco-rated switches, such as the ones mentioned above as Boundary Clocks (T-BC); plus a FibroLAN Falcon-RX or a Qulsar QG2 as  GrandMaster Clocks (T-GM). Both T-GM are connected to an external GPS antenna through a coaxial cable and a splitter.
Additionally, all Dell servers are equipped with SmartNICs (Intel E810) to accelerate RAN applications.

\noindent\textbf{Programmable Logical Topology:}
We leverage automated configuration of 802.1q \glspl{vlan} \cite{ieee99vlan} to support multiple logical topologies within the same physical environment, enabling {\bf R1, R2}, and {\bf R5}.
VLANs add a 32-bit field to the Ethernet frame, with 12 reserved for the VLAN ID (VID), thus allowing for up to 4096 different \glspl{vlan} to co-exist. 
Programmable switches then uses the the VID to forward (or drop) the packets at each port accordingly.

\begin{figure}
    \centering
    \includegraphics[width=1\linewidth]{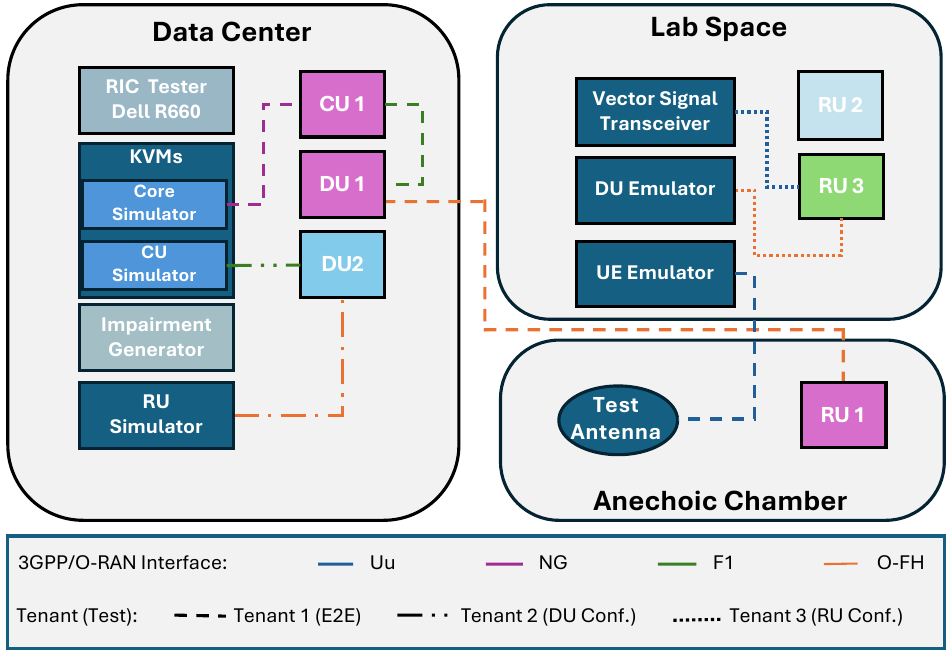}
    
    \caption{Logical topology representing three different tenants running tests in parallel. Tenant 1 is running a conducted \gls{e2e} test on CU1 and DU1, Tenant 2 is running WG4 conformance for \gls{du}2, and Tenant 3 is running a WG4 conformance test for \gls{ru}1 (radiated).}
    \label{fig:logical}
\end{figure}

\glspl{vlan} can be configured manually on the switch. However, the complexity of the OTIC deployment and testing requirements calls for automated solutions to programmatically configure the VLANs, supported by most switches through built-in \glspl{api} or \gls{sdn} controllers.
Fig.~\ref{fig:logical} shows how the logical infrastructure allow multiple test cases to be run in parallel, with different colors representing different 3GPP and O-RAN interfaces, and different line types (dashed, dotted, dash dotted) represents RU conformance, DU conformance, and E2E tests. Each combination of line type and color represents a different \gls{vlan}.

\noindent\textbf{Network Allocation and Segmentation:} A careful design of the network from a L2 and L3 point of view is required to address {\bf R1, R2, R4, R5}.
As detailed in Table \ref{tab:otic_network}, an IPv4 /16 network is reserved for the OTIC operations, with three /24 networks (\gls{oob}, management, and services) allocated for internal usage. 
Then, up to 100 individual /24 networks (254 addresses each) are available to be assigned to tenants. Finally, a number of additional subnets is reserved for the communication among different tenants, each one specific to a different 5G interface.
In the same table we can also see how only certain subnets are routed in the \gls{otic} infrastructure, allowing them to be reached from the VPN or from other hosts in the network.

\begin{table}[t]
    \centering
    \begin{tabular}{llll}
         Type & Name & Subnet & Routable \\
         \midrule
         \multirowcell{3}[1ex][l]{OTIC Networks} & \gls{oob} & X.Y.0.0/24 & Y\\
          & Management & X.Y.1.0/24 & Y\\
          & Services & X.Y.2.0/24 & Y\\
         \midrule
        \multirowcell{6}[1ex][l]{Tenant Networks} &Tenant 1 & X.Y.4.0/24 & Y\\
         &Tenant 2& X.Y.5.0/24 & Y\\
         & $\dots$ & $\dots$\\
        \midrule
        \multirowcell{6}[1ex][l]{5G Data Networks}
        &F1 & X.Y.101.0/24 & N\\
        &NG & X.Y.102.0/24 & N\\
        &O1 & X.Y.103.0/24 & N\\
        &E1 & X.Y.104.0/24 & N\\
        &\gls{ofh} M-plane & X.Y.105.0/24 & Y\\
        &\gls{ofh} CU-Plane & L2 only &N\\
        & $\dots$ & $\dots$\\
    \end{tabular}
    \caption{Network segmentation of a /16 IPv4 subnet into OTIC Internal Networks, Tenant specific Networks and 5G Data Networks.}
    \vspace{-.3cm}
    \label{tab:otic_network}
\end{table}

\begin{table}[t]
    \centering
    \begin{tabular}{lll}
         Type & Name & Subnet \\
         \midrule
         \multirowcell{3}[1ex][l]{\gls{te} Tenant 1} & Management & X.Y.4.0/26 \\
          & OOB & X.Y.4.128/27 \\
          & VPN & X.Y.4.160/29\\
          & $\dots$ & $\dots$\\
          \midrule
          \multirowcell{3}[1ex][l]{\gls{dut} Tenant 2} & Management & X.Y.10.0/26 \\
          & OOB & X.Y.10.128/27 \\
          & VPN & X.Y.10.160/29\\
          & $\dots$ & $\dots$\\
    \end{tabular}
    \caption{Segmentation of two tenants specific networks.}
    \label{tab:tenant_network}
\end{table}

The tenants' subnets are then subdivided into smaller ones according to the individual needs of each tenant. Table \ref{tab:tenant_network} provides an overview of the most common networks, among which we find Management, \gls{oob}, and a subnet for \gls{vpn} access.
To ensure isolation between tenants, each tenant access is restricted only to its own /24 subnet and the OTIC services hosted in the dedicated subnet.

5G Data Networks are also subdivided into smaller subnets. A larger one (/26) is used for shared access between tenants (e.g., during a Plugfest). Multiple smaller ones (/29) are tailored for specific tests and mapped to specific \gls{otic} tenants. 
As an example, Table \ref{tab:5g_networks} shows a shared 5G data interface (i.e., NG, used for the communication between \glspl{cu} and the \gls{amf} service in the 5G Core) in the /26 shared subnet. Smaller subnets, mapped to specific \glspl{vlan}, are automatically configured only to support a specific testing activity and associated connectivity across \glspl{dut} and \glspl{te}.

\begin{table}[]
    \centering
    \begin{tabular}{lll}
         Type & Name & Subnet \\
        \midrule
         \multirowcell{3}[1ex][l]{NG}
        &Shared & X.Y.102.0/26 \\
        &E2E Tenants 2-3 & X.Y.102.64/29 \\
        &WG4 Conf. Tenant 1 & X.Y.102.72/29 \\
        & $\dots$ & $\dots$\\
    \end{tabular}
    \caption{Segmentation of one 5G Data networks into networks specific for single/multiple tenants testing activities.}
        \vspace{-.3cm}
    \label{tab:5g_networks}
\end{table}

\noindent\textbf{Services:}
The OTIC infrastructure requires a series of services to support the testing activities and requirements discussed in Sec.~\ref{sec:infra-requirements}. A directory server (e.g., based on LDAP) manages identities and authentication across the infrastructure. 
A VPN appliance provides secure remote access ({\bf R5}) and border peering with upstream connectivity providers. In our case, the VPN is also used to connect to other testbeds, e.g., the X5G setup which is deployed in the Northeastern Boston campus, meeting requirement {\bf R7}.

A DNS server separates hostnames for services from IPs for more flexibility and resiliency in configuring the system ({\bf R4}). 
Storage is provided through NFS and a network attached storage unit. 
The inventory system is used to maintain a list of all the equipment with locations in all the different sites that the OTIC operates. It is also used to track physical connections among servers, as well as IP allocation. For this, we leverage Netbox.\footnote{\url{https://github.com/netbox-community/netbox}} For monitoring,  we use Observium,\footnote{\url{https://www.observium.org/}} to monitor active ports and traffic using the SNMP protocol, and Base Management Controllers (BMCs) to remotely access the servers using the \gls{oob} networks. We also leverage Nagios to monitor services uptime and system stability.

\section{Conclusions}
\label{sec:conclusions}

In this paper, we introduced a blueprint for the design and deployment of a programmable \gls{otic} that can address challenges in the O-RAN and 3GPP testing space. We reviewed the testing options within the O-RAN specifications, including conformance, interoperability, and end-to-end testing. We then highlight eight key requirements for the OTIC infrastructure, and presented a blueprint that provides components and solutions that meet such requirements. The blueprint is based on a physical topology that can be programmed and instrumented to support multiple tests, even from different tenants, in isolation, without the need for manual reconfiguration of the system. We believe that this discussion can guide the development of O-RAN testing practices. As future work, we will further extend the infrastructure to support automated RIC testing and additional automation workflows.

\bibliographystyle{IEEEtran}
\bibliography{otic_blueprint.bib}
\end{document}

%% file: acronyms.tex
\newacronym{3gpp}{3GPP}{3rd Generation Partnership Project}
\newacronym{4g}{4G}{4th generation}
\newacronym{5g}{5G}{5th generation}
\newacronym{5gc}{5GC}{5G Core}
\newacronym{adc}{ADC}{Analog to Digital Converter}
\newacronym{aerpaw}{AERPAW}{Aerial Experimentation and Research Platform for Advanced Wireless}
\newacronym{ai}{AI}{Artificial Intelligence}
\newacronym{aimd}{AIMD}{Additive Increase Multiplicative Decrease}
\newacronym{am}{AM}{Acknowledged Mode}
\newacronym{amc}{AMC}{Adaptive Modulation and Coding}
\newacronym{amf}{AMF}{Access and Mobility Management Function}
\newacronym{aops}{AOPS}{Adaptive Order Prediction Scheduling}
\newacronym{api}{API}{Application Programming Interface}
\newacronym{apn}{APN}{Access Point Name}
\newacronym{aqm}{AQM}{Active Queue Management}
\newacronym{ausf}{AUSF}{Authentication Server Function}
\newacronym{avc}{AVC}{Advanced Video Coding}
\newacronym{awgn}{AGWN}{Additive White Gaussian Noise}
\newacronym{balia}{BALIA}{Balanced Link Adaptation Algorithm}
\newacronym{bbu}{BBU}{Base Band Unit}
\newacronym{bdp}{BDP}{Bandwidth-Delay Product}
\newacronym{ber}{BER}{Bit Error Rate}
\newacronym{bf}{BF}{Beamforming}
\newacronym{bler}{BLER}{Block Error Rate}
\newacronym{brr}{BRR}{Bayesian Ridge Regressor}
\newacronym{bsr}{BSR}{Buffer Status Report}
\newacronym{bs}{BS}{Base Station}
\newacronym{bss}{BSS}{Business Support System}
\newacronym{ca}{CA}{Carrier Aggregation}
\newacronym{caas}{CaaS}{Connectivity-as-a-Service}
\newacronym{cb}{CB}{Code Block}
\newacronym{cc}{CC}{Congestion Control}
\newacronym{ccid}{CCID}{Congestion Control ID}
\newacronym{cco}{CC}{Carrier Component}
\newacronym{cdd}{CDD}{Cyclic Delay Diversity}
\newacronym{cdf}{CDF}{Cumulative Distribution Function}
\newacronym{cdn}{CDN}{Content Distribution Network}
\newacronym{cir}{CIR}{Channel Impulse Response}
\newacronym{cn}{CN}{Core Network}
\newacronym{codel}{CoDel}{Controlled Delay Management}
\newacronym{comac}{COMAC}{Converged Multi-Access and Core}
\newacronym{cord}{CORD}{Central Office Re-architected as a Datacenter}
\newacronym{cornet}{CORNET}{COgnitive Radio NETwork}
\newacronym{cosmos}{COSMOS}{Cloud Enhanced Open Software Defined Mobile Wireless Testbed for City-Scale Deployment}
\newacronym{cots}{COTS}{Commercial Off-the-Shelf}
\newacronym{cp}{CP}{Control Plane}
\newacronym{cpu}{CPU}{Central Processing Unit}
\newacronym{cqi}{CQI}{Channel Quality Information}
\newacronym{cr}{CR}{Cognitive Radio}
\newacronym{cran}{CRAN}{Cloud \gls{ran}}
\newacronym{crs}{CRS}{Cell Reference Signal}
\newacronym{csi}{CSI}{Channel State Information}
\newacronym{csirs}{CSI-RS}{Channel State Information - Reference Signal}
\newacronym{cu}{CU}{Central Unit}
\newacronym{d2tcp}{D$^2$TCP}{Deadline-aware Data center TCP}
\newacronym{d3}{D$^3$}{Deadline-Driven Delivery}
\newacronym{dac}{DAC}{Digital to Analog Converter}
\newacronym{dag}{DAG}{Directed Acyclic Graph}
\newacronym{darpa}{DARPA}{Defense Advanced Research Projects Agency}
\newacronym{das}{DAS}{Distributed Antenna System}
\newacronym{dash}{DASH}{Dynamic Adaptive Streaming over HTTP}
\newacronym{dc}{DC}{Dual Connectivity}
\newacronym{dccp}{DCCP}{Datagram Congestion Control Protocol}
\newacronym{dce}{DCE}{Direct Code Execution}
\newacronym{ssbs}{SSBs}{synchronization signal blocks}
\newacronym{ebs}{EBS}{exhaustive beam sweep}
\newacronym{dci}{DCI}{Downlink Control Information}
\newacronym{dcl}{DCL}{Dear Colleague Letter}
\newacronym{dctcp}{DCTCP}{Data Center TCP}
\newacronym{dl}{DL}{Downlink}
\newacronym{dmr}{DMR}{Deadline Miss Ratio}
\newacronym{drlcc}{DRL-CC}{Deep Reinforcement Learning Congestion Control}
\newacronym{drs}{DRS}{Discovery Reference Signal}
\newacronym{du}{DU}{Distributed Unit}
\newacronym{ecaas}{ECaaS}{Edge-Cloud-as-a-Service}
\newacronym{ecn}{ECN}{Explicit Congestion Notification}
\newacronym{edf}{EDF}{Earliest Deadline First}
\newacronym{isi}{ISI}{inter-stream interference}
\newacronym{embb}{eMBB}{Enhanced Mobile Broadband}
\newacronym{empower}{EMPOWER}{EMpowering transatlantic PlatfOrms for advanced WirEless Research}
\newacronym{enb}{eNB}{evolved Node Base}
\newacronym{endc}{EN-DC}{E-UTRAN-\gls{nr} \gls{dc}}
\newacronym{epc}{EPC}{Evolved Packet Core}
\newacronym{eps}{EPS}{Evolved Packet System}
\newacronym{es}{ES}{Edge Server}
\newacronym{etsi}{ETSI}{European Telecommunications Standards Institute}
\newacronym[firstplural=Estimated Times of Arrival (ETAs)]{eta}{ETA}{Estimated Time of Arrival}
\newacronym{eutran}{E-UTRAN}{Evolved Universal Terrestrial Access Network}
\newacronym{faas}{FaaS}{Function-as-a-Service}
\newacronym{fapi}{FAPI}{Functional Application Platform Interface}
\newacronym{fcc}{FCC}{Federal Communications Commission}
\newacronym{fdd}{FDD}{Frequency Division Duplexing}
\newacronym{fdm}{FDM}{Frequency Division Multiplexing}
\newacronym{fdma}{FDMA}{Frequency Division Multiple Access}
\newacronym{fed4fire}{FED4FIRE+}{Federation 4 Future Internet Research and Experimentation Plus}
\newacronym{fir}{FIR}{Finite Impulse Response}
\newacronym{fit}{FIT}{Future \acrlong{iot}}
\newacronym{fpga}{FPGA}{Field Programmable Gate Array}
\newacronym{fr2}{FR2}{Frequency Range 2}
\newacronym{fs}{FS}{Fast Switching}
\newacronym{fscc}{FSCC}{Flow Sharing Congestion Control}
\newacronym{ftp}{FTP}{File Transfer Protocol}
\newacronym{fw}{FW}{Flow Window}
\newacronym{ge}{GE}{Gaussian Elimination}
\newacronym{gnb}{gNB}{Next Generation Node Base}
\newacronym{gop}{GOP}{Group of Pictures}
\newacronym{gpr}{GPR}{Gaussian Process Regressor}
\newacronym{gpu}{GPU}{Graphics Processing Unit}
\newacronym{gtp}{GTP}{GPRS Tunneling Protocol}
\newacronym{gtpc}{GTP-C}{GPRS Tunnelling Protocol Control Plane}
\newacronym{gtpu}{GTP-U}{GPRS Tunnelling Protocol User Plane}
\newacronym{gtpv2c}{GTPv2-C}{\gls{gtp} v2 - Control}
\newacronym{gw}{GW}{Gateway}
\newacronym{harq}{HARQ}{Hybrid Automatic Repeat reQuest}
\newacronym{hetnet}{HetNet}{Heterogeneous Network}
\newacronym{hh}{HH}{Hard Handover}
\newacronym{hol}{HOL}{Head-of-Line}
\newacronym{hqf}{HQF}{Highest-quality-first}
\newacronym{hss}{HSS}{Home Subscription Server}
\newacronym{http}{HTTP}{HyperText Transfer Protocol}
\newacronym{ia}{IA}{Initial Access}
\newacronym{iab}{IAB}{Integrated Access and Backhaul}
\newacronym{ic}{IC}{Incident Command}
\newacronym{ietf}{IETF}{Internet Engineering Task Force}
\newacronym{imsi}{IMSI}{International Mobile Subscriber Identity}
\newacronym{imt}{IMT}{International Mobile Telecommunication}
\newacronym{ip}{IP}{Internet Protocol}
\newacronym{itu}{ITU}{International Telecommunication Union}
\newacronym{kpi}{KPI}{Key Performance Indicator}
\newacronym{kvm}{KVM}{Kernel-based Virtual Machine}
\newacronym{los}{LOS}{Line-of-Sight}
\newacronym{lsm}{LSM}{Link-to-System Mapping}
\newacronym{lstm}{LSTM}{Long Short Term Memory}
\newacronym{lte}{LTE}{Long Term Evolution}
\newacronym{lxc}{LXC}{Linux Container}
\newacronym{m2m}{M2M}{Machine to Machine}
\newacronym{mac}{MAC}{Medium Access Control}
\newacronym{manet}{MANET}{Mobile Ad Hoc Network}
\newacronym{mano}{MANO}{Management and Orchestration}
\newacronym{mc}{MC}{Multi-Connectivity}
\newacronym{mcc}{MCC}{Mobile Cloud Computing}
\newacronym{mchem}{MCHEM}{Massive Channel Emulator}
\newacronym{mcs}{MCS}{Modulation and Coding Scheme}
\newacronym{mec}{MEC}{Multi-access Edge Computing}
\newacronym{mec2}{MEC}{Mobile Edge Cloud}
\newacronym{mfc}{MFC}{Mobile Fog Computing}
\newacronym{mi}{MI}{Mutual Information}
\newacronym{mib}{MIB}{Master Information Block}
\newacronym{miesm}{MIESM}{Mutual Information Based Effective SINR}
\newacronym{mimo}{MIMO}{Multiple Input, Multiple Output}
\newacronym{m-mimo}{Massive MIMO}{Massive Multiple Input, Multiple Output}
\newacronym{mgen}{MGEN}{Multi-Generator}
\newacronym{ml}{ML}{Machine Learning}
\newacronym{mlr}{MLR}{Maximum-local-rate}
\newacronym[plural=\gls{mme}s,firstplural=Mobility Management Entities (MMEs)]{mme}{MME}{Mobility Management Entity}
\newacronym{mmtc}{mMTC}{Massive Machine-Type Communications}
\newacronym{mmwave}{mmWave}{millimeter wave}
\newacronym{mpdccp}{MP-DCCP}{Multipath Datagram Congestion Control Protocol}
\newacronym{mptcp}{MPTCP}{Multipath TCP}
\newacronym{mr}{MR}{Maximum Rate}
\newacronym{mrdc}{MR-DC}{Multi \gls{rat} \gls{dc}}
\newacronym{mse}{MSE}{Mean Square Error}
\newacronym{mss}{MSS}{Maximum Segment Size}
\newacronym{mt}{MT}{Mobile Termination}
\newacronym{mtd}{MTD}{Machine-Type Device}
\newacronym{mtu}{MTU}{Maximum Transmission Unit}
\newacronym{mumimo}{MU-MIMO}{Multi-user \gls{mimo}}
\newacronym{mvno}{MVNO}{Mobile Virtual Network Operator}
\newacronym{nalu}{NALU}{Network Abstraction Layer Unit}
\newacronym{nas}{NAS}{Network Attached Storage}
\newacronym{nbiot}{NB-IoT}{Narrow Band IoT}
\newacronym{nfv}{NFV}{Network Function Virtualization}
\newacronym{nfvi}{NFVI}{Network Function Virtualization Infrastructure}
\newacronym{nic}{NIC}{Network Interface Card}
\newacronym{nlos}{NLOS}{Non-Line-of-Sight}
\newacronym{now}{NOW}{Non Overlapping Window}
\newacronym{nrdz}{NRDZ}{National Radio Dynamic Zone}
\newacronym{nsf}{NSF}{National Science Foundation}
\newacronym{nsm}{NSM}{Network Service Mesh}
\newacronym[type=hidden]{nr}{NR}{New Radio}
\newacronym{nrf}{NRF}{Network Repository Function}
\newacronym{nsa}{NSA}{Non Stand Alone}
\newacronym{nse}{NSE}{Network Slicing Engine}
\newacronym{nssf}{NSSF}{Network Slice Selection Function}
\newacronym{ntp}{NTP}{Network Time Protocol}
\newacronym{o2i}{O2I}{Outdoor to Indoor}
\newacronym{oai}{OAI}{OpenAirInterface}
\newacronym{oaif}{OAIF}{OpenAirInterface Foundation}
\newacronym{oaicn}{OAI-CN}{\gls{oai} \acrlong{cn}}
\newacronym{oairan}{OAI-RAN}{\acrlong{oai} \acrlong{ran}}
\newacronym{oam}{OAM}{Operations, Administration and Maintenance}
\newacronym{ofdm}{OFDM}{Orthogonal Frequency Division Multiplexing}
\newacronym{olia}{OLIA}{Opportunistic Linked Increase Algorithm}
\newacronym{omec}{OMEC}{Open Mobile Evolved Core}
\newacronym{onap}{ONAP}{Open Network Automation Platform}
\newacronym{onf}{ONF}{Open Networking Foundation}
\newacronym{onos}{ONOS}{Open Networking Operating System}
\newacronym{oom}{OOM}{\gls{onap} Operations Manager}
\newacronym{opnfv}{OPNFV}{Open Platform for \gls{nfv}}
\newacronym[type=hidden]{oran}{O-RAN}{Open \gls{ran}}
\newacronym{orbit}{ORBIT}{Open-Access Research Testbed for Next-Generation Wireless Networks}
\newacronym{os}{OS}{Operating System}
\newacronym{oss}{OSS}{Operations Support System}
\newacronym{pa}{PA}{Position-aware}
\newacronym{pase}{PASE}{Prioritization, Arbitration, and Self-adjusting Endpoints}
\newacronym{pawr}{PAWR}{Platforms for Advanced Wireless Research}
\newacronym{pbch}{PBCH}{Physical Broadcast Channel}
\newacronym{pcef}{PCEF}{Policy and Charging Enforcement Function}
\newacronym{pcfich}{PCFICH}{Physical Control Format Indicator Channel}
\newacronym{pcrf}{PCRF}{Policy and Charging Rules Function}
\newacronym{pdcch}{PDCCH}{Physical Downlink Control Channel}
\newacronym{pdcp}{PDCP}{Packet Data Convergence Protocol}
\newacronym{pdsch}{PDSCH}{Physical Downlink Shared Channel}
\newacronym{pdu}{PDU}{Packet Data Unit}
\newacronym{pf}{PF}{Proportional Fair}
\newacronym{pgw}{PGW}{Packet Gateway}
\newacronym{phich}{PHICH}{Physical Hybrid ARQ Indicator Channel}
\newacronym{phy}{PHY}{Physical}
\newacronym{pmch}{PMCH}{Physical Multicast Channel}
\newacronym{pmi}{PMI}{Precoding Matrix Indicators}
\newacronym{powder}{POWDER}{Platform for Open Wireless Data-driven Experimental Research}
\newacronym{ppo}{PPO}{Proximal Policy Optimization}
\newacronym{ppp}{PPP}{Poisson Point Process}
\newacronym{prach}{PRACH}{Physical Random Access Channel}
\newacronym{prb}{PRB}{Physical Resource Block}
\newacronym{psnr}{PSNR}{Peak Signal to Noise Ratio}
\newacronym{pss}{PSS}{Primary Synchronization Signal}
\newacronym{pucch}{PUCCH}{Physical Uplink Control Channel}
\newacronym{pusch}{PUSCH}{Physical Uplink Shared Channel}
\newacronym{qam}{QAM}{Quadrature Amplitude Modulation}
\newacronym{qci}{QCI}{\gls{qos} Class Identifier}
\newacronym{qoe}{QoE}{Quality of Experience}
\newacronym{qos}{QoS}{Quality of Service}
\newacronym{quic}{QUIC}{Quick UDP Internet Connections}
\newacronym{rach}{RACH}{Random Access Channel}
\newacronym[plural=Radio Access Networks (RAN)]{ran}{RAN}{Radio Access Network}
\newacronym[firstplural=Radio Access Technologies (RATs)]{rat}{RAT}{Radio Access Technology}
\newacronym{rcn}{RCN}{Research Coordination Network}
\newacronym{rec}{REC}{Radio Edge Cloud}
\newacronym{red}{RED}{Random Early Detection}
\newacronym{renew}{RENEW}{Reconfigurable Eco-system for Next-generation End-to-end Wireless}
\newacronym{rf}{RF}{Radio Frequency}
\newacronym{rfc}{RFC}{Request for Comments}
\newacronym{rfr}{RFR}{Random Forest Regressor}
\newacronym{ric}{RIC}{RAN Intelligent Controller}
\newacronym{rlc}{RLC}{Radio Link Control}
\newacronym{rlf}{RLF}{Radio Link Failure}
\newacronym{rlnc}{RLNC}{Random Linear Network Coding}
\newacronym{rmse}{RMSE}{Root Mean Squared Error}
\newacronym{rnis}{RNIS}{Radio Network Information Service}
\newacronym{rr}{RR}{Round Robin}
\newacronym{rrc}{RRC}{Radio Resource Control}
\newacronym{rrm}{RRM}{Radio Resource Management}
\newacronym{rru}{RRU}{Remote Radio Unit}
\newacronym{rs}{RS}{Remote Server}
\newacronym{rsrp}{RSRP}{Reference Signal Received Power}
\newacronym{rsrq}{RSRQ}{Reference Signal Received Quality}
\newacronym{rss}{RSS}{Received Signal Strength}
\newacronym{rssi}{RSSI}{Received Signal Strength Indicator}
\newacronym{rtt}{RTT}{Round Trip Time}
\newacronym{ru}{RU}{Radio Unit}
\newacronym{rw}{RW}{Receive Window}
\newacronym{rx}{RX}{Receiver}
\newacronym{s1ap}{S1AP}{S1 Application Protocol}
\newacronym{sa}{SA}{standalone}
\newacronym{sack}{SACK}{Selective Acknowledgment}
\newacronym{sap}{SAP}{Service Access Point}
\newacronym{sc2}{SC2}{Spectrum Collaboration Challenge}
\newacronym{scef}{SCEF}{Service Capability Exposure Function}
\newacronym{sch}{SCH}{Secondary Cell Handover}
\newacronym{scoot}{SCOOT}{Split Cycle Offset Optimization Technique}
\newacronym{sctp}{SCTP}{Stream Control Transmission Protocol}
\newacronym{sdap}{SDAP}{Service Data Adaptation Protocol}
\newacronym{sdk}{SDK}{Software Development Kit}
\newacronym{sdm}{SDM}{Space Division Multiplexing}
\newacronym{sdma}{SDMA}{Spatial Division Multiple Access}
\newacronym{sdn}{SDN}{Software-defined Networking}
\newacronym{sdr}{SDR}{Software-defined Radio}
\newacronym{seba}{SEBA}{SDN-Enabled Broadband Access}
\newacronym{sgsn}{SGSN}{Serving GPRS Support Node}
\newacronym{sgw}{SGW}{Service Gateway}
\newacronym{si}{SI}{Study Item}
\newacronym{sib}{SIB}{Secondary Information Block}
\newacronym{sinr}{SINR}{Signal to Interference plus Noise Ratio}
\newacronym{sip}{SIP}{Session Initiation Protocol}
\newacronym{siso}{SISO}{Single Input, Single Output}
\newacronym{sla}{SLA}{Service Level Agreement}
\newacronym{sm}{SM}{Saturation Mode}
\newacronym{smf}{SMF}{Session Management Function}
\newacronym{smo}{SMO}{Service Management and Orchestration}
\newacronym{sms}{SMS}{Short Message Service}
\newacronym{smsgmsc}{SMS-GMSC}{\gls{sms}-Gateway}
\newacronym{snr}{SNR}{Signal-to-Noise-Ratio}
\newacronym{son}{SON}{Self-Organizing Network}
\newacronym{sptcp}{SPTCP}{Single Path TCP}
\newacronym{srb}{SRB}{Service Radio Bearer}
\newacronym{srn}{SRN}{Standard Radio Node}
\newacronym{srs}{SRS}{Sounding Reference Signal}
\newacronym{ss}{SS}{Synchronization Signal}
\newacronym{sss}{SSS}{Secondary Synchronization Signal}
\newacronym{st}{ST}{Spanning Tree}
\newacronym{svc}{SVC}{Scalable Video Coding}
\newacronym{tb}{TB}{Transport Block}
\newacronym{tcp}{TCP}{Transmission Control Protocol}
\newacronym{tdd}{TDD}{Time Division Duplexing}
\newacronym{tdm}{TDM}{Time Division Multiplexing}
\newacronym{tdma}{TDMA}{Time Division Multiple Access}
\newacronym{tfl}{TfL}{Transport for London}
\newacronym{tfrc}{TFRC}{TCP-Friendly Rate Control}
\newacronym{tft}{TFT}{Traffic Flow Template}
\newacronym{tgen}{TGEN}{Traffic Generator}
\newacronym{tip}{TIP}{Telecom Infra Project}
\newacronym{tm}{TM}{Transparent Mode}
\newacronym{to}{TO}{Telco Operator}
\newacronym{tr}{TR}{Technical Report}
\newacronym{trp}{TRP}{Transmitter Receiver Pair}
\newacronym{ts}{TS}{Technical Specification}
\newacronym{tti}{TTI}{Transmission Time Interval}
\newacronym{ttt}{TTT}{Time-to-Trigger}
\newacronym{tx}{TX}{Transmitter}
\newacronym{uas}{UAS}{Unmanned Aerial System}
\newacronym{uav}{UAV}{Unmanned Aerial Vehicle}
\newacronym{udm}{UDM}{Unified Data Management}
\newacronym{udp}{UDP}{User Datagram Protocol}
\newacronym{udr}{UDR}{Unified Data Repository}
\newacronym{ue}{UE}{User Equipment}
\newacronym{uhd}{UHD}{\gls{usrp} Hardware Driver}
\newacronym{ul}{UL}{Uplink}
\newacronym{um}{UM}{Unacknowledged Mode}
\newacronym{uml}{UML}{Unified Modeling Language}
\newacronym{upa}{UPA}{Uniform Planar Array}
\newacronym{upf}{UPF}{User Plane Function}
\newacronym{urllc}{URLLC}{Ultra Reliable and Low Latency Communication}
\newacronym{usa}{U.S.}{United States}
\newacronym{usim}{USIM}{Universal Subscriber Identity Module}
\newacronym{usrp}{USRP}{Universal Software Radio Peripheral}
\newacronym{utc}{UTC}{Urban Traffic Control}
\newacronym{vim}{VIM}{Virtualization Infrastructure Manager}
\newacronym{vm}{VM}{Virtual Machine}
\newacronym{vnf}{VNF}{Virtual Network Function}
\newacronym{volte}{VoLTE}{Voice over \gls{lte}}
\newacronym{voltha}{VOLTHA}{Virtual OLT HArdware Abstraction}
\newacronym{vr}{VR}{Virtual Reality}
\newacronym{vran}{vRAN}{Virtualized \gls{ran}}
\newacronym{vss}{VSS}{Video Streaming Server}
\newacronym{wbf}{WBF}{Wired Bias Function}
\newacronym{wf}{WF}{Wired-first}
\newacronym{wlan}{WLAN}{Wireless Local Area Network}
\newacronym{osm}{OSM}{Open Source \gls{nfv} Management and Orchestration}
\newacronym{pnf}{PNF}{Physical Network Function}
\newacronym{drl}{DRL}{Deep Reinforcement Learning}
\newacronym{mtc}{MTC}{Machine-type Communications}

\newacronym{cif}{CI}{cyberinfrastructure}
\newacronym{sonic}{SONiC}{Software for Open Networking in the Cloud}
\newacronym{ocp}{OCP}{Open Compute Project}
\newacronym{snmp}{SNMP}{Simple Network Management Protocol}
\newacronym{raid}{RAID}{redundant array of independent disks}
\newacronym{nfs}{NFS}{Network File Storage}
\newacronym{ci}{CI}{Continuous Integration}
\newacronym{cd}{CD}{Continuous Deployment}
\newacronym{dtn}{DTN}{Data Transfer Node}

\newacronym{dns}{DNS}{Domain Name Service}
\newacronym{nrpe}{NRPE}{Nagios Remote Plugin Executor}
\newacronym{ldap}{LDAP}{Lightweight Directory Access Protocol}
\newacronym{lan}{LAN}{Local Area Network}
\newacronym{vlan}{VLAN}{Virtual LAN}

\newacronym{ipmi}{IPMI}{Intelligent Platform Management Interface}
\newacronym{tor}{ToR}{Top-of-the-Rack}
\newacronym{lmn}{LMN}{Large Memory Node}
\newacronym{bgp}{BGP}{Border Gateway Protocol}
\newacronym{dhcp}{DHCP}{Dynamic Host Configuration Protocol}
\newacronym{vrf}{VRF}{Virtual Routing and Forwarding}
\newacronym{vpn}{VPN}{Virtual Private Network}
\newacronym{rma}{RMA}{Return Merchandise Authorization}
\newacronym{hpc}{HPC}{High Performance Compute}

\newacronym{nu}{NU}{Northeastern University}
\newacronym{asic}{ASIC}{Application-specific Integrated Circuit}
\newacronym{rdma}{RDMA}{Remote Direct Memory Access}
\newacronym{roce}{RoCE}{RDMA over Converged Ethernet}
\newacronym{ovs}{OVS}{Open vSwitch}
\newacronym{frr}{FRR}{Free Range Routing}
\newacronym{ups}{UPS}{Uninterruptible Power Supply}

\newacronym{ntia}{NTIA}{National Telecommunications and Information Administration}
\newacronym{pii}{PII}{Personal and Identifiable Information}
\newacronym{irb}{IRB}{Institutional Review Board}
\newacronym{doi}{DOI}{Digital Object Identifier}

\newacronym{sdo}{SDO}{Standard-Development Organization}
\newacronym{ose}{OSE}{Open Source Ecosystem}
\newacronym{osc}{OSC}{O-RAN Software Community}
\newacronym{dop}{DOP}{Director of Operations}
\newacronym{pm}{PM}{Program Manager}
\newacronym{excom}{EXCOM}{Executive Committee}
\newacronym{iiot}{IIoT}{Industrial \gls{iot}}
\newacronym{lf}{LF}{Linux Foundation}

\newacronym{wiot}{WIoT}{Institute for the Wireless Internet of Things}
\newacronym{nyu}{NYU}{New York University}
\newacronym{otic}{OTIC}{Open Testing \& Integration Center}

\newacronym{nofo}{NOFO}{Notice of Funding Opportunity}

\newacronym{onr}{ONR}{Office of Naval Research}
\newacronym{afosr}{AFOSR}{Air Force Office of Scientific Research}
\newacronym{afrl}{AFRL}{Air Force Research Laboratory}
\newacronym{arl}{ARL}{Army Research Laboratory}

\newacronym{arc}{ARC}{Aerial Research Cloud}

\newacronym{mno}{MNO}{Mobile Network Operator}
\newacronym{ct}{CT}{Continuous Testing}
\newacronym{oci}{OCI}{Open Container Initiative}
\newacronym{scf}{SCF}{Small Cell Forum}
\newacronym{dsp}{DSP}{Digital Signal Processing}
\newacronym{ulpi}{ULPI}{Uplink Performance Improvement}
\newacronym{abf}{ABF}{Antenna \& Beamforming}
\newacronym{iot-test}{IOT}{Interoperability Test}
\newacronym{dmrs}{DMRS}{Demodulation Reference Signal}
\newacronym{aoa}{AoA}{Angle of Arrival}
\newacronym[firstplural=Devices Under Test (DUTs)]{dut}{DUT}{Device Under Test}
\newacronym{sut}{SUT}{System Under Test}
\newacronym{ofh}{O-FH}{Open Fronthaul}
\newacronym{wat}{WAT}{Wraparound Testing}
\newacronym{tifg}{TIFG}{Testing and Integration Focus Group}
\newacronym{iot}{IOT}{Interoperability and Testing}
\newacronym{e2e}{E2E}{End to End}
\newacronym{te}{TE}{Testing Equipment}
\newacronym{oob}{OOB}{Out of Band}
\newacronym{vst}{VST}{Vector Signal Transceiver}
\newacronym{lacp}{LACP}{Link Aggregation Control Protocol}
\newacronym{wg}{WG}{Working Group}

%% file: otic_blueprint.bbl
\begin{thebibliography}{10}
\providecommand{\url}[1]{#1}
\csname url@samestyle\endcsname
\providecommand{\newblock}{\relax}
\providecommand{\bibinfo}[2]{#2}
\providecommand{\BIBentrySTDinterwordspacing}{\spaceskip=0pt\relax}
\providecommand{\BIBentryALTinterwordstretchfactor}{4}
\providecommand{\BIBentryALTinterwordspacing}{\spaceskip=\fontdimen2\font plus
\BIBentryALTinterwordstretchfactor\fontdimen3\font minus
  \fontdimen4\font\relax}
\providecommand{\BIBforeignlanguage}[2]{{%
\expandafter\ifx\csname l@#1\endcsname\relax
\typeout{** WARNING: IEEEtran.bst: No hyphenation pattern has been}%
\typeout{** loaded for the language `#1'. Using the pattern for}%
\typeout{** the default language instead.}%
\else
\language=\csname l@#1\endcsname
\fi
#2}}
\providecommand{\BIBdecl}{\relax}
\BIBdecl

\bibitem{polese2023understanding}
M.~Polese, L.~Bonati \emph{et~al.}, ``{Understanding O-RAN: Architecture,
  Interfaces, Algorithms, Security, and Research Challenges},'' \emph{IEEE
  Communications Surveys \& Tutorials}, vol.~25, no.~2, pp. 1376--1411, Second
  quarter 2023.

\bibitem{dellOroRAN}
\BIBentryALTinterwordspacing
S.~Pongratz, ``{Open RAN Market Opportunity and Risks},'' Dell'Oro Group
  report, 2021. [Online]. Available:
  \url{https://www.delloro.com/wp-content/uploads/2021/11/DellOro-Group-Article-Open-RAN-Market-Opportunity-and-Risks.pdf}
\BIBentrySTDinterwordspacing

\bibitem{bala2021ric}
B.~Balasubramanian, E.~S. Daniels \emph{et~al.}, ``{RIC: A RAN Intelligent
  Controller Platform for AI-Enabled Cellular Networks},'' \emph{IEEE Internet
  Computing}, vol.~25, no.~2, pp. 7--17, March 2021.

\bibitem{groen2024securing}
J.~Groen, S.~D'Oro \emph{et~al.}, ``{Securing O-RAN Open Interfaces},''
  \emph{IEEE Transactions on Mobile Computing}, pp. 1--13, 2024.

\bibitem{klement2024securing}
F.~Klement, A.~Brighente \emph{et~al.}, ``{Securing the Open RAN
  Infrastructure: Exploring Vulnerabilities in Kubernetes Deployments},'' in
  \emph{IEEE 10th International Conference on Network Softwarization
  (NetSoft)}, June 2024, pp. 185--189.

\bibitem{garciaaviles2021nuberu}
G.~Garcia-Aviles, A.~Garcia-Saavedra \emph{et~al.}, ``Nuberu: Reliable {RAN}
  virtualization in shared platforms,'' in \emph{Proceedings of ACM MobiCom},
  New Orleans, LA, USA, October 2021.

\bibitem{cagr_private}
{Grand View Research}, ``{Private 5G Network Market Size, Share \& Trends
  Analysis Report},''
  \url{https://www.grandviewresearch.com/industry-analysis/private-5g-network-market}.

\bibitem{brown2023heavy}
\BIBentryALTinterwordspacing
G.~Brown, ``{Heavy Reading’s 2023 Open RAN Operator Survey},'' 2023.
  [Online]. Available:
  \url{https://img.lightreading.com/downloads/whitepapers/HR-2023-Open-RAN-Operator-Survey.pdf}
\BIBentrySTDinterwordspacing

\bibitem{oranWebsiteOticPage}
\BIBentryALTinterwordspacing
{O-RAN ALLIANCE}, ``{Open Testing \& Integration Centres (OTIC)},'' 2024.
  [Online]. Available: \url{https://www.o-ran.org/testing-integration}
\BIBentrySTDinterwordspacing

\bibitem{oran23overview}
{O-RAN Alliance}, ``{Overview of Open Testing and Integration Centre (OTIC) and
  O-RAN Certification and Badging Program},'' \emph{White Paper}, 2023.

\bibitem{oran23wg4conf}
{Working Group 4}, ``{Conformance Test Specification},'' {O-RAN Alliance},
  Technical Specification (TS), 2023, revision 03.00.09.

\bibitem{oran23cgotic}
{Test and Integration Focus Group}, ``{Criteria and Guidelines of Open Testing
  and Integration Centre},'' {O-RAN Alliance}, Technical Specification (TS),
  2023, revision 05.00.

\bibitem{oran23wg4iot}
{Working Group 4}, ``{Fronthaul Interoperability Test Specification},'' {O-RAN
  Alliance}, Technical Specification (TS), 2022, revision 03.00.10.

\bibitem{oran23tifge2e}
{Testing and Integration Focus Group}, ``{End-to-end Test Specification},''
  {O-RAN Alliance}, Technical Specification (TS), 2022, revision 04.00.

\bibitem{oran23wg5iot}
{Working Group 5}, ``{Fronthaul Interoperability Test Specification},'' {O-RAN
  Alliance}, Technical Specification (TS), 2022, revision 10.

\bibitem{villa2024x5gopenprogrammablemultivendor}
\BIBentryALTinterwordspacing
D.~Villa, I.~Khan \emph{et~al.}, ``{X5G: An Open, Programmable, Multi-vendor,
  End-to-end, Private 5G O-RAN Testbed with NVIDIA ARC and OpenAirInterface},''
  2024. [Online]. Available: \url{https://arxiv.org/abs/2406.15935}
\BIBentrySTDinterwordspacing

\bibitem{polese2024colosseumopenrandigital}
\BIBentryALTinterwordspacing
M.~Polese, L.~Bonati \emph{et~al.}, ``{Colosseum: The Open RAN Digital Twin},''
  2024. [Online]. Available: \url{https://arxiv.org/abs/2404.17317}
\BIBentrySTDinterwordspacing

\bibitem{ieee99vlan}
``{IEEE Standards for Local and Metropolitan Area Networks: Virtual Bridged
  Local Area Networks},'' \emph{{IEEE Std 802.1Q-1998}}, pp. 1--214, 1999.

\end{thebibliography}
